\documentclass{article}

\usepackage[nonatbib,final]{neurips_2021}

\usepackage[utf8]{inputenc} %
\usepackage[T1]{fontenc}    %
\usepackage{hyperref}       %
\usepackage{url}            %
\usepackage{booktabs}       %
\usepackage{amsfonts}       %
\usepackage{nicefrac}       %
\usepackage{microtype}      %
\usepackage{xcolor}         %
\usepackage{multirow}
\usepackage{graphicx}
\usepackage{amsmath}
\usepackage{graphbox}
\usepackage{caption}

\usepackage[caption=false]{subfig}

\renewcommand{\vec}[1]{\mathbf{#1}}
\title{Towards Learned Simulators for Cell Migration}

\author{%
  Koen Minartz \\
  Eindhoven University of Technology\\
  \texttt{k.minartz@tue.nl} \\
    \And
   Yoeri Poels \\
   Eindhoven University of Technology \\
   \texttt{y.r.j.poels@tue.nl} \\
   \And
   Vlado Menkovski \\
   Eindhoven University of Technology \\
   \texttt{v.menkovski@tue.nl} 
}

\begin{document}
\maketitle

\begin{abstract}

Simulators driven by deep learning are gaining popularity as a tool for efficiently emulating accurate but expensive numerical simulators. Successful applications of such \emph{neural} simulators can be found in the domains of physics, chemistry, and structural biology, amongst others. Likewise, a neural simulator for cellular dynamics can augment lab experiments and traditional computational methods to enhance our understanding of a cell's interaction with its physical environment. In this work, we propose an autoregressive probabilistic model that can reproduce spatiotemporal dynamics of single cell migration, traditionally simulated with the Cellular Potts model. We observe that standard single-step training methods do not only lead to inconsistent rollout stability, but also fail to accurately capture the stochastic aspects of the dynamics, and we propose training strategies to mitigate these issues. Our evaluation on two proof-of-concept experimental scenarios shows that neural methods have the potential to faithfully simulate stochastic cellular dynamics at least an order of magnitude faster than a state-of-the-art implementation of the Cellular Potts model.

\end{abstract}

\section{Introduction}

Studying the variety of mechanisms through which cells migrate and interact with their physical environment is of crucial importance for our understanding of cell biology. For example, cell migration plays a key role in the interaction between the immune system and implant surfaces~\cite{Dalton2001, Thenard2020}, the development of embryos~\cite{Scarpa2016}, and the progression of cancer~\cite{Spatarelu2019, Kumar2016}. As experimental capacity in the lab is inherently limited, computational methods have emerged as a tool to investigate the stochastic and dynamic movement and shape of cells. However, these methods can be computationally demanding. This is especially restrictive in scenarios requiring many simulations, for example due to substantial stochasticity or in the case of inverse design, where parameters are optimized iteratively based on the simulator's output. Moreover, parameterizing such models to realistically simulate cells can be a difficult task, requiring careful design and expert knowledge.

On the other hand, deep learning has been gaining traction as a tool for learning fast approximate simulators. %
For example, for continuous-time and continuous-space systems defined with partial differential equations (PDEs), neural solvers learn to emulate a system's dynamics from a dataset of simulations generated by a more computationally demanding solver~\cite{fno, stachenfeld2021learned, brandstetter2022message, iakovlev2021, wu2022learning}. In this setting, a large computational cost is paid up front to generate the training set, but once trained, approximate solutions can be generated at a fraction of the original cost. Moreover, learned simulators hold the promise of emulating systems for which the laws governing the dynamics are not known, by instead training on experimental observations.

Based on the above considerations, we propose to use neural simulators to simulate cellular dynamics. More specifically, we consider the scenario where both the movement and shape of the cell show stochastic aspects and are highly dynamic, which is typically modeled in the Cellular Potts modeling framework, proposed in~\cite{Graner1992}. Given their various successful applications in modeling spatiotemporal data, we hypothesize that neural simulators are capable of faithfully emulating the ground truth dynamics, while  accelerating the simulation process. Our contributions are summarized as follows:

\begin{itemize}
    \item We propose a neural simulation model to simulate stochastic single-cell dynamics similar to those generated by the Cellular Potts model;
    \item We develop and evaluate autoregressive training strategies, with the aim to improve the model's rollout performance and its ability to capture stochastic dynamics;
    \item We observe that our method has the capacity to faithfully emulate the cellular dynamics of the Cellular Potts model, while generating simulations an order of magnitude faster.
\end{itemize}

\section{Background and Related Work}\label{sec:background-related-work}
\subsection{Cellular Potts Model}\label{sec:cp-model}

The Cellular Potts (CP) model is a computational modeling framework for simulating cellular dynamics and the dynamic and fluctuating morphology of cells on a lattice~\cite{Graner1992, Savill1997, Balter2007}. %
The CP model has gained prominence due to its flexibility in modeling cell shape and movement, the interaction between multiple cells, stochastic aspects of cell behavior, and multiscale mechanisms~\cite{Scianna2012, Rens2019, Hirashima2017}.

 In the CP framework, the system is modeled as a Euclidean lattice $L$ and Hamiltonian $H$. The function $x: L \rightarrow S$ maps each lattice site $l_i \in L$ to its state $x(l_i) \in S$, where $S$ is the set of all cells and materials that can be present in the system. Note that in the CP literature $x$ is commonly referred to as $\sigma$; we deviate from this to stick to machine learning convention. %
 To evolve the system, a Markov-Chain Monte Carlo sampling algorithm is used. At every iteration, a lattice site $l_i$ is chosen at random. Then, a proposal is made to modify $x$ such that state $x(l_i)$ is changed to $x(l_j)$, where $l_j$ is a site adjacent to $l_i$. Finally, the difference in energy $\Delta H$ is calculated between the proposed and current system state. If $\Delta H \leq 0$, the proposed state is accepted as the new system state; if $\Delta H > 0$, it is accepted with probability $e^{\frac{- \Delta H}{T}}$, with $T$ being the \emph{temperature} parameter of the model.

 The Hamiltonian $H$ itself differs per application, but typically consists of at least contact energy and volume preservation terms, as originally proposed in~\cite{Graner1992}:
\begin{equation}\label{eq:hamiltonian-glazier}
    H = \underbrace{\sum_{l_i,l_j \in \mathcal{N}(L)} J\left(x(l_i), x(l_j) \right) \left(1-\delta_{x(l_i), x(l_j)}\right)}_{\text{contact energy}} + \underbrace{\sum_{c \in C} \lambda_V \left(V(c) - V^*(c)\right)^2}_{\text{volume preservation}} + H_{\text{other}},
\end{equation}
where $\mathcal{N}(L)$ is the set of all pairs of neighboring lattice sites in $L$,  $J\left(x(l_i), x(l_j)\right)$ is the contact energy between cells and/or materials $x(l_i)$ and $x(l_j)$, and $\delta_{x, y}$ is the Kronecker delta. Furthermore, $C$ is the set of all cells in the system, $V(c)$ is the number of lattice sites occupied by cell $c$ (from here on referred to as the \emph{volume} of cell $c$), $V^*(c)$ is the target volume of cell $c$, and $\lambda_V$ is a Lagrange multiplier. $H_\text{other}$ can consist of many extensions and modifications of the original Hamiltonian, for example taking into account cellular dynamics induced by forces, gradients in chemical concentrations, cell surface area constraints, and many more biological concepts. The specific Hamiltonians used for simulating our data can be found in Appendix~\ref{app:hamiltonians}.

\subsection{Neural Simulators}
Neural networks have been employed for simulation in many domains~\cite{molecularsim2020,fluidmechanics2020}, often by either combining ML models with existing numerical solvers~\cite{um2020,kochkov2021} or by using ML models to simulate dynamics in their entirety~\cite{brandstetter2022message,fno,stachenfeld2021learned}. The latter, which we refer to as \textit{neural simulators}, encompass the type of model proposed in this work, as we seek to emulate the CP simulations as a whole. Of particular interest are autoregressive methods operating on a spatial grid, as these fit both the temporal and spatial component of the CP simulations. This setup generally comes with challenges of ensuring prediction quality and stability over longer rollout trajectories. Common approaches to address this include injecting noise and incorporating model rollouts in the training procedure~\cite{stachenfeld2021learned, brandstetter2022message}.

In the context of cellular dynamics, neural networks have been used in various settings to aid in simulation. TrajectoryNet~\cite{tong2020} utilizes optimal transport in combination with continuous normalizing flows to interpolate time-evolving gene expression data based on population measurements at fixed timepoints. In a similar spirit, CellBox~\cite{yuan2021, ji2021} simulates molecular biological processes with neural ODEs~\cite{chen2018}. LEUP~\cite{Barua2020} models cells as atomic particles moving in two-dimensional space to investigate collective cell migration. However, these methods do not address simulation of spatiotemporal dynamics of cells on a grid. Another branch of related research focuses on grid-based generative modeling of cell morphology and subcellular organization~\cite{Donovan-Maiye2022, Goldsborough2017, Ruan2018}. Very recently, Wiesner et al.~\cite{wiesner2022} proposed a method to model time evolving cell shapes. Although these methods can impressively perform (conditional) cell image generation, they do not address migration dynamics.

\section{Method}\label{sec:approach}
 
\subsection{Machine Learning Formulation}\label{sec:ML-formulation}

\begin{figure}[!b]
    \centering
    \includegraphics[width=1.0\linewidth, align=c]{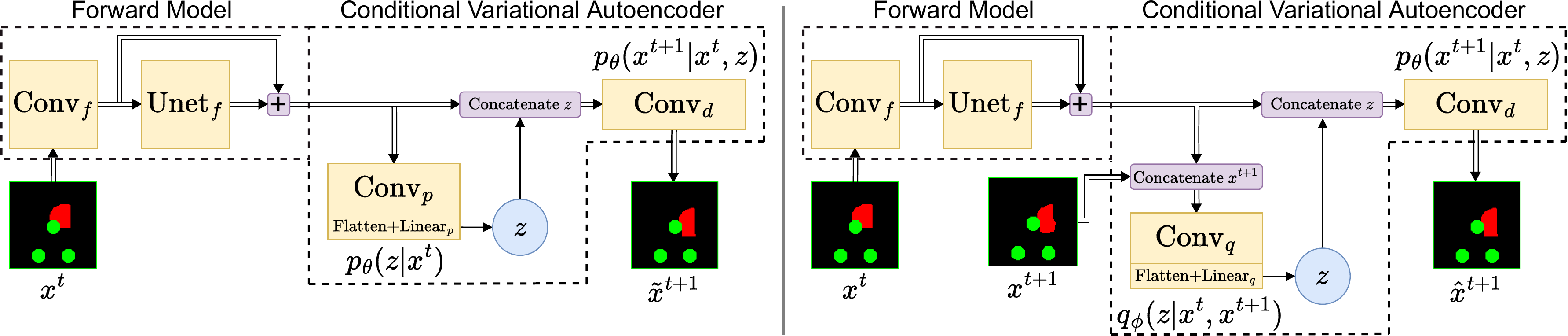}%
    \caption{Illustrations of the model's generative procedure (left) and inference procedure (right). Note that the geometry of the system is maintained throughout the model, as indicated by the double line.}
    \label{fig:models}
\end{figure}
We consider the problem of learning an autoregressive probabilistic mapping to simulate spatiotemporal cellular dynamics. More formally, at each time $t$ the system is described by its state $x^t$, a categorically-valued function on a fixed grid (see also Section~\ref{sec:cp-model}). Correspondingly, a system evolution is specified by a sequence of states $x^{0:T}$. We postulate a ground-truth probability distribution $p^*(x^{0:T})$ over system evolutions $x^{0:T}$ from which we can sample using the CP simulator. Hence, our aim is to learn the parameters $\theta$ of the model $p_\theta$ such that
\begin{equation}
\mathbb{E}_{x^{0:T} \sim p^*} \log p_{\theta}(x^{0:T})
\end{equation}
is maximized. By the Markov property of the data generating process of the CP model configurations that we consider, and with $p^*(x^0)$ a specified distribution of initial conditions, this is equivalent to maximizing Equation $\ref{eq:opt_obj}$ with respect to $\theta$:
\begin{equation}\label{eq:opt_obj}
\mathbb{E}_{t \sim U\{0, T-1\}} \mathbb{E}_{x^{t} \sim p^*}  \log p_{\theta}(x^{t+1} | x^{t}).
\end{equation}
Consequently, we aim to model $p^*(x^{0:T})$ by learning $p_{\theta}(x^{t+1} | x^{t})$ and applying it autoregressively.

\subsection{Model Design}\label{sec:modeldesign}

The design choices for modeling $p_{\theta}(x^{t+1} | x^{t})$ are driven by the goals to produce a model that is capable of producing realistic trajectories with good sample efficiency.
To achieve high sample efficiency, we introduce latent variable $z$, following the Conditional Variational Autoencoder (CVAE) framework~\cite{Sohn2015}. In our case, $z$ is conditioned on previous state $x^{t}$ according to conditional prior $p_\theta(z|x^t)$ and the subsequent state $x^{t+1}$ follows the distribution $p_\theta(x^{t+1}|x^t,z)$.

As our model is an autoregressive model, it consists of a \textit{forward model} that evolves the state of the system to the next state. However, rather than directly producing the pixel-wise parameters of a distribution to sample the next state from, the representation produced by the forward model is used to first condition the prior distribution of the latent variable. Then, to produce a sample of the next state, we sample the latent variable from this prior and combine it with the forward representation, which is subsequently decoded. This procedure is depicted in Figure~\ref{fig:models} (left). 

To enable the model to produce realistic trajectories, we align the model's structure with that of the data. Consequently, we maintain the geometry of the system in the forward representations, which in practice means using a fully convolutional architecture in the forward model as well as the decoder. Specifically, the forward model is implemented using the U-net architecture~\cite{Ronneberger2015}, but it could in principle be substituted for any other fully convolutional architecture. Additionally, we do not compress all information in a latent space without geometrical structure, but allow the decoder to access the intermediate, geometrically meaningful representation produced by the forward model. 

The use of a U-net is further motivated by the desire to model dynamics on various spatial scales. Note that, as mentioned before, latent variable $z$ does not share the system geometry. To sidestep this issue, $z$ is broadcasted to the entire domain before being concatenated to the internal representation, as done in a probabilistic U-net~\cite{Kohl2018}. Further architectural details are provided in Appendix~\ref{app:model-arch}.

\subsection{Training and Rollout Stability}\label{sec:training-rollout}
As per the VAE framework~\cite{Kingma2014}, an approximate posterior distribution $q_\phi(z|x^{t},x^{t+1})$ is learned, also referred to as the inference network. Note that this distribution is inferred from both the current and next timestep. During training, we optimize the Evidence Lower Bound (ELBO) on the log-likelihood:
\begin{equation}\label{eq:elbo}
\log p_{\theta}(x^{t+1} | x^{t}) \geq -KL(q_\phi(z|x^{t},x^{t+1})||p_\theta(z|x^t)) + \mathbb{E}_{q_\phi(z|x^{t},x^{t+1})}[\log p_\theta(x^{t+1}|x^t,z)].
\end{equation}
The inference procedure is depicted in Figure~\ref{fig:models} (right). During trajectory generation we sample latent variable $z$ from the prior distribution $p_\theta(z|x^t)$. By then sampling $\tilde{x}^{t+1} \sim p_\theta(x^{t+1}|x^t, z)$ and repeating the entire process, we can simulate longer trajectories.

One of the major challenges of autoregressive neural simulators is the instability of long rollouts. As the model is applied iteratively, errors accumulate, causing $p_\theta(x^{t + r} | x^t)$ to stray from the data distribution for sufficiently large rollout length $r \in \mathbb{N}^+$. To mitigate this issue, various methods have been proposed, typically involving some form of noise injection or rollout training. We consider two training methods. 

The \textbf{Multi-step} training procedure applies the model for $r$ iterations and calculates the loss for each iteration. At each step $t$, a reconstruction $\hat{x}^{t+1}$ is created by sampling from $p_\theta(\hat{x}^{t+1} | x^t, z) q_\phi (z | x^t, x^{t+1})$. $\hat{x}^{t+1}$ then serves as the next input for the model, which is used to create a reconstruction $\hat{x}^{t+2}$, and so on. Note that, since $\hat{x}^{t+1}$ is discrete, the gradients are only backpropagated a single step. The \textbf{Pushforward} training method, adapted from~\cite{brandstetter2022message}, is similar in nature. We sample $r \sim U[1, r_{\max}]$, rollout the model $r$ times as described for multi-step training, but now only calculate and backpropagate the loss at the final iteration. In this way, the model learns to correct its own rollout errors and to map back to the data distribution. From a probabilistic perspective, the rollout errors can be seen as adding noise to the data, where the noise is sampled from the model's error distribution itself.

Orthogonal to the training procedure, we also consider two ways to sample from $p_\theta$ to improve stability. Both have in common that $z$ is sampled from the posterior (when training) or the prior (when generating new simulations), but differ in how $\tilde{x}^{t+1}$ is sampled from $p_\theta(x^{t+1} |z, x^t)$. With \textbf{maximum likelihood} sampling, a single $z$ is sampled, and $\tilde{x}^{t+1}$ is discretized such that $p_\theta(x^{t+1} | z, x^t)$ is maximized for that $z$. However, this method does not exploit any domain knowledge about the system. With \textbf{volume preservation} sampling, knowledge about the (approximately) preserved volume of the cell is injected in the method. Here, $\tilde{x}^{t+1}$ is also discretized to maximize $p_\theta(x^{t+1} | z, x^t)$, but under the additional constraint that the volume of the cell equals its target volume.

\section{Experiments}\label{sec:results}

Our evaluation consists of two experiments. In the first, randomly scattered walls are placed on a grid, and a force pointing to the bottom center of the lattice is applied to the cell. Despite the comparatively simple setting, local stochasticity in the system already leads to interesting behavior such as cell shape fluctuations, and even emerging bifurcations of trajectories.%

The second experiment consists of a system where more global stochasticity with bifurcating trajectories is simulated. A downwards force is applied to the cell, such that it makes contact with pillars placed at fixed positions along its path. Then, either a leftward or rightward force is applied to the cell with equal probability, such that it passes each pillar on the left or right. The system always has approximately identical initial conditions, but a strongly multi-modal distribution over trajectories.

All Cellular Potts simulations are generated with CompuCell3D~\cite{Swat2012}, the state-of-the-art software package for CP model simulations. We implement python extensions in order to generate the desired dynamics. In all experiments, one time unit corresponds to 500 Monte Carlo steps in CompuCell3D. For each experiment, we use 1350 training, 150 validation and 150 testing trajectories. 

We compare maximum likelihood and volume preservation sampling and investigate various training strategies, see Section~\ref{sec:training-rollout} for details on both. All models are trained for 300 epochs. We optimize the ELBO with the reparameterization trick~\cite{Kingma2014} and use a linear KL-annealing schedule~\cite{Bowman2016}. We set the maximum rollout length $r_{\max}$ to a third of the total trajectory length for pushforward and multi-step training.

\subsection{Simple Dynamics}\label{sec:simple-dyn-exp}

Quantitatively, we evaluate models with two metrics: the estimated ELBO on the log-likelihood (LL), and the \emph{unrolled reconstruction log-likelihood} (URLL). LL is calculated as the sum of the lower bounds on the one-step conditional log-likelihoods (see also Equations~\ref{eq:opt_obj} and~\ref{eq:elbo}). URLL is a heuristic metric for assessing rollout stability. %
Starting from initial state $x^0$, we estimate $p_\theta(x^1 | x^0)$, and at each subsequent step, the log-likelihood of $x^t$ is estimated from $p_\theta(x^t |\hat{x}^{t-1})$, where $\hat{x}^{t-1} \sim p_\theta(\hat{x}^{t-1} | \hat{x}^{t-2}, z) q_\phi(z|\hat{x}^{t-2}, x^{t-1})$ is a sampled \emph{reconstruction} of the model at the preceding timestep. A high URLL should correspond to a model that can reconstruct entire trajectories well, rather than only individual steps.
\begin{table}[t]
\centering
\begin{tabular}{cc|cc}
Training strategy               & Sampling method               & LL &  URLL \\ \hline
\multirow{2}{*}{One-step}       & Maximum likelihood & $\geq$ -3211.9               & $\geq$ -40669.8       \\
                                & Volume preservation & $\geq$ -4049.3            & $\geq$ -100243.9      \\ \hline
\multirow{2}{*}{Pushforward}    & Maximum likelihood & $\geq$ -6973.3                 & $\geq$ -36920.8        \\
                                & Volume preservation & $\geq$ -3874.3            & $\geq$ -12975.8             \\ \hline
\multirow{2}{*}{Multi-step}     & Maximum likelihood & $\geq$ \textbf{-2751.0}                 & $\geq$ \textbf{-4619.9}       \\
                                & Volume preservation & $\geq$ -3031.5            & $\geq$ -4689.1                  
\end{tabular}
\caption{Evaluation of training and sampling strategy combinations (best values marked in bold).}
\label{tab:exp-1}
\end{table}

\begin{figure}[b]
    \centering
    \includegraphics[width=0.46\linewidth]{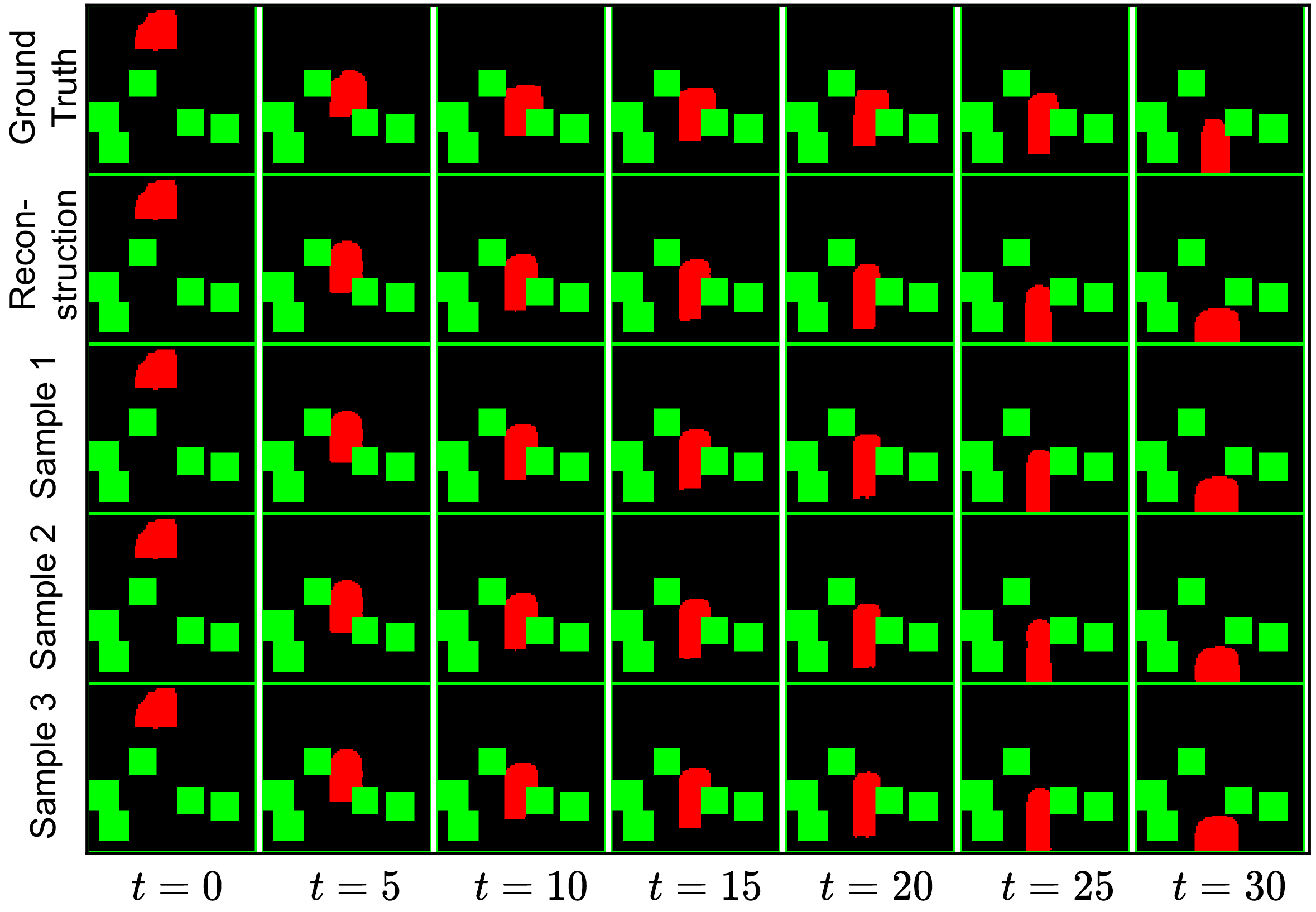}\hspace{.02\linewidth}
    \includegraphics[width=0.46\linewidth]{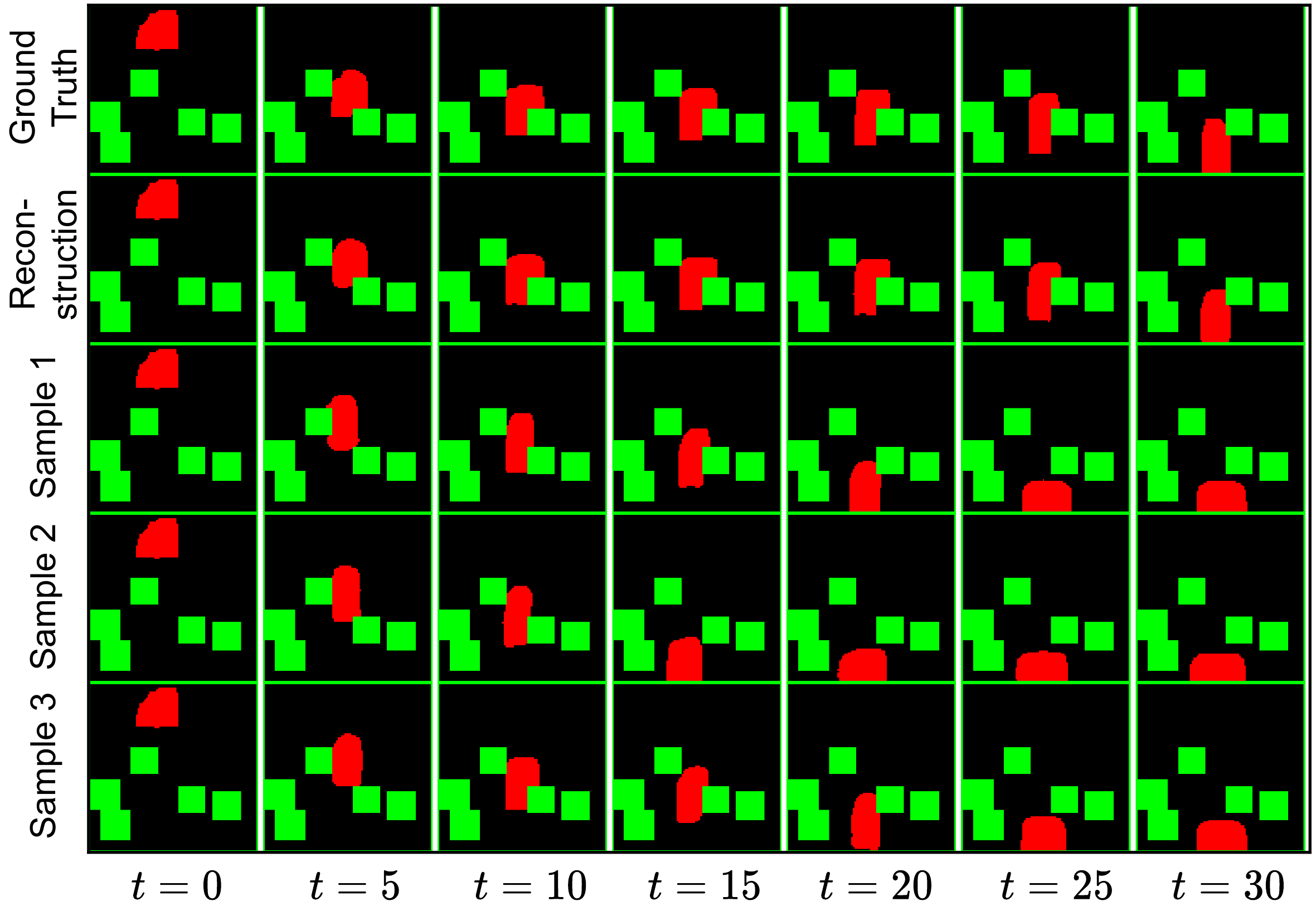}
    \caption{Ground truth and sampled trajectories, using volume preservation sampling and one-step (left) and multi-step (right) training. The cell is depicted in red, walls are depicted in green.}
    \label{fig:model-rollouts}
\end{figure}

The results of simulators with varying training and sampling strategies are given in Table~\ref{tab:exp-1}. As expected, training strategies such as pushforward and multi-step training improve the URLL score for both sampling methods, compared to single-step training. Interestingly, pushforward training is not always beneficial for single-step log-likelihoods, whereas multi-step training does show an improvement. Another interesting result is that volume preservation sampling does not necessarily improve upon maximum likelihood sampling. We expected that the former would net an advantage as domain knowledge is integrated in the sampling procedure, but the results are not conclusive. In fact, multi-step training with maximum likelihood sampling performs best for both metrics, albeit with a small margin.

Qualitatively, the simulators can capture the stochastic behavior of the system, both locally in the form of a fluctuating cell membrane, and globally, generating diverse but realistic trajectories from identical initial conditions. However, this behavior is not captured well when using single-step training. This is exemplified in Figure~\ref{fig:model-rollouts}, which shows reconstructed and sampled trajectories for models using volume preservation, trained with one-step and multi-step training. %
Despite the probabilistic approach, one-step training results in samples that are almost identical. In contrast, the same architecture trained with multi-step training is capable of generating a variety of realistic trajectories.

We also measured the time required for one model rollout of 60 steps (corresponding to 30000 Monte Carlo steps for the CP model simulation) on commodity hardware (details in Appendix~\ref{app:hardware}). Over 100 repetitions, the average model rollout time was $2.16$ seconds ($\sigma=0.08s$) on our CPU, and $0.56$ seconds ($\sigma=0.01s$) on our GPU. Generating a simulation with CompuCell3D took $14.16$ seconds on average ($\sigma = 0.76$) on the same CPU, and GPU acceleration is not supported by CompuCell3D for simulations not involving chemical diffusion. Consequently, even for these simple dynamics, the neural simulator already provides an 85\% speedup compared to the CP model on identical hardware. Moreover, we envision speedups to become even larger when the dynamics are more involved, for example, dynamics involving diffusion of chemicals or multi-cellular systems.

\subsection{Bifurcating dynamics}\label{sec:bifur-dyn-exp}
\begin{figure}[t]
\begin{minipage}[b]{0.529\textwidth}
    \centering
    \includegraphics[width=0.9\linewidth]{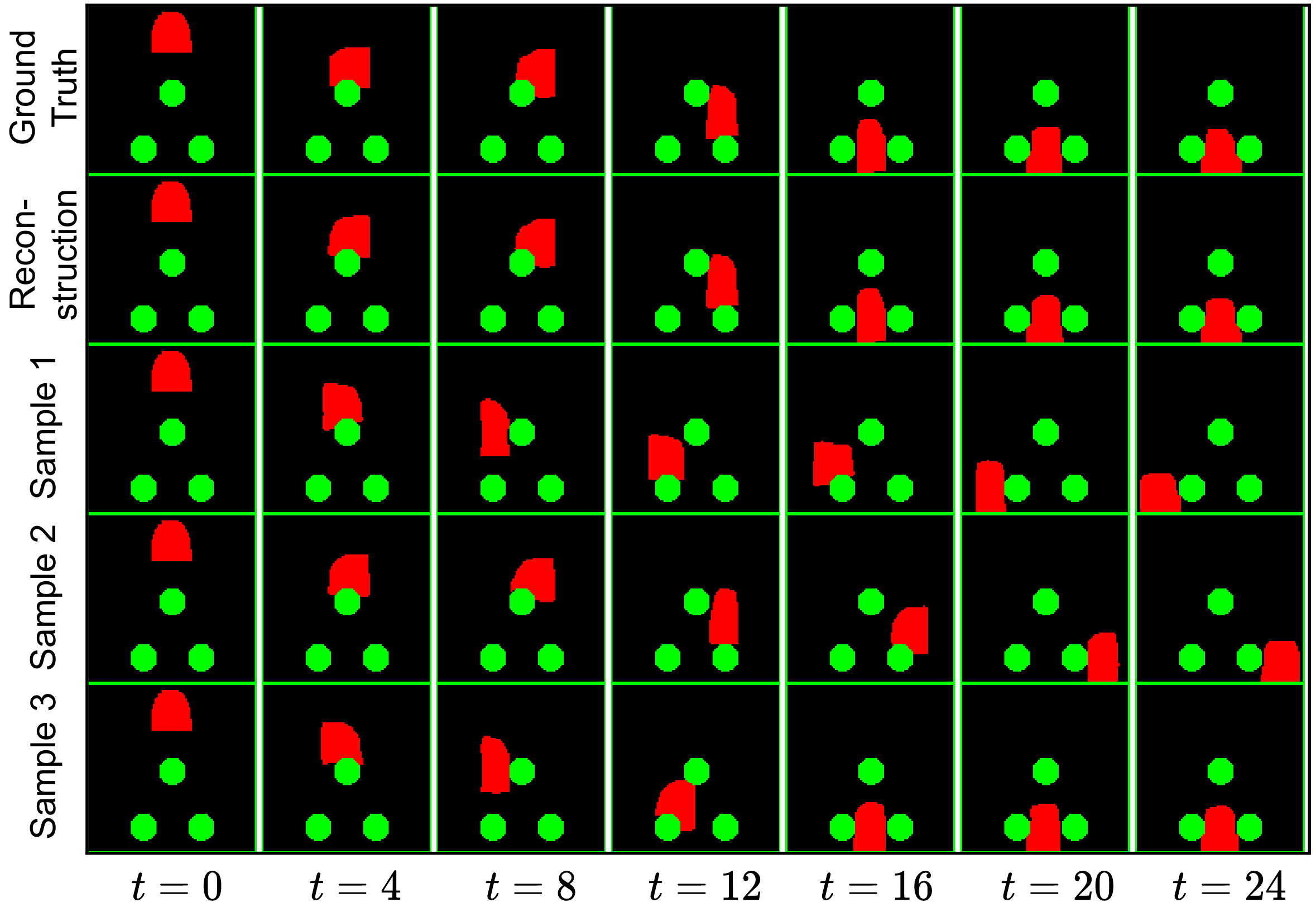}%
    \captionof{figure}{Trajectories generated by the CP model and the volume-preserving model with multi-step training.
    }
    \label{fig:bifurcation}
\end{minipage}
\hfill
\begin{minipage}[b]{0.45\textwidth}
\centering
\setlength{\tabcolsep}{4pt}
\begin{tabular}{c|ccc}
Training method       & Left & Center & Right \\ \hline
One-step    & 0.27 & 0.72   & 0.01     \\
Pushforward  & 0.03 & 0.54   & 0.43  \\
Multi-step   & 0.25 & 0.52   & 0.23  \\ \hline
Cellular Potts       & 0.23 & 0.5   & 0.27  \\ \hline
Expected & 0.25 & 0.5    & 0.25 
\end{tabular}
\captionof{table}{Empirical distribution of the x-coordinate of the cell's center of mass at the end of the simulation. Trajectories are sampled from the volume-preserving model trained with various training strategies, and from the Cellular Potts model.}
\label{tab:bifur-hist}
\end{minipage}
\end{figure}
Figure~\ref{fig:bifurcation} shows trajectories for bifurcating dynamics with identical initial conditions. Observe that the model generates trajectories in which the cell follows varying realistic paths. To quantitatively assess the training strategies, we examine the position of the cell at the end of the simulation. If the neural simulator works well, these aggregate statistics should match those of the trajectories generated by the CP model. Here, we present volume-preserving models as we found them to work better in this scenario; results for models using maximum likelihood sampling are provided in Appendix~\ref{app:mlbif}. The end positions' distributions are found in Table~\ref{tab:bifur-hist}. The expected distribution closely matches that of the CP model and our method trained with multi-step training. However, when trained with one-step or pushforward training, the distribution of the neural simulator differs substantially from the true distribution. This demonstrates that a probabilistic model in itself is not sufficient to capture global, long-term stochastic aspects of the evolution of the system, and training methods that go beyond single-step predictions are necessary. 

\section{Conclusion}\label{sec:conclusion}

In this work, we proposed a probabilistic neural simulation model for spatiotemporal cellular dynamics. We adapted training strategies from autoregressive models and found that these improve rollout quality and enable the model to accurately capture the stochastic dynamics of the system. To evaluate our method, we generated data using the CP model and show that the learned simulator is capable of faithfully emulating the dynamics. %
Furthermore, sampling from the learned simulator is around an order of magnitude faster than the CP simulations that generated the training data. We conclude that neural simulators are a promising method for simulating spatiotemporal cellular dynamics, with many interesting avenues for relevant research, for example simulating systems with multiple cells, or integrating neural cellular dynamics simulators and neural PDE solvers for multiscale modeling.

\bibliographystyle{plain}
\bibliography{main}
\newpage
\appendix

\section{Hamiltonians for Simple and Bifurcating Dynamics}\label{app:hamiltonians}

The Hamiltonian for both simple and bifurcating dynamics (Sections~\ref{sec:simple-dyn-exp} and~\ref{sec:bifur-dyn-exp}) is as follows:
\begin{align*}
    H &= \underbrace{\sum_{l_i,l_j \in \mathcal{N}(L)} J\left(x(l_i), x(l_j) \right) \left(1-\delta_{x(l_i), x(l_j)}\right)}_\text{contact energy}\\
    &+  \underbrace{\lambda_v \left(V(c) - V^*(c)\right)^2}_\text{volume constraint} \\
    &+ \underbrace{\lambda_a \left(A(c) - A^*(c)\right)^2}_\text{surface area constraint} \\
    &+ \underbrace{\vec{\lambda_F}^T \cdot \vec{COM(c)}}_\text{external potential}, 
\end{align*}
where $\vec{COM(c)}$ is the center-of-mass vector of the cell $c$.

The values for each of the parameters for both simple and bifurcating dynamics are given in Table~\ref{tab:app-cp-params}:

\begin{table}[h]
\centering
\setlength{\tabcolsep}{4pt}
\begin{tabular}{c|cccccccc}
            & $T$ & $J(\text{cell}, \text{medium})$ & $J(\text{cell}, \text{wall})$ & $\lambda_V$ & $V^*(c)$ & $\lambda_A$ & $A^*(c)$ & $\vec{\lambda_F}$             \\ \hline
Simple      & 15  & 8                               & 10                            & 5           & 500       & 1           & 70       & $[F_x^\text{simple}, -35]^T$    \\
Bifurcating & 15  & 8                               & 16                            & 5           & 500       & 1           & 70       & $[F_x^\text{bifurcate}, -35]^T$
\end{tabular}
\caption{CP parameters for both experiments}
\label{tab:app-cp-params}
\end{table}

Here, $F_x^\text{simple} = \frac{50 - COM(c)_x}{50} \cdot 35$ is continuously updated throughout the CP simulation. For the bifurcating dynamics, if the cell is not in contact with a micropillar or if the cell reached the bottom of the grid, $F_x^\text{bifurcate} = 0$. If the cell comes into contact with a micropillar and $F_x^\text{bifurcate}$ equals 0, it is set to a value with magnitude equal to the contact area with a micropillar, of which the sign is negative or positive with equal probability. Finally, if the cell is still in contact with a micropillar and $F_x^\text{bifurcate}$ is nonzero, the horizontal force is rescaled such that its magnitude equals the contact area of the micropillar. We explicitly note that these parameters were not necessarily chosen to be biologically plausible, but to generate stylized CP simulations that exhibit behavior which will be relevant for modeling more complex, realistic scenarios as well.

\section{Model Architecture Details}\label{app:model-arch}

The details of all model components are given below. All convolutional layers are 2D convolutions with kernel size 3x3, unless otherwise mentioned. 

\begin{itemize}
    \item \textbf{Forward Model:} The forward model starts with a linear convolutional layer to lift the input to a higher dimensional representation. This representation is processed by a U-net. The U-net has 4 contrastive blocks, one block that operates on the lowest level of the U-net, and 4 expansive blocks. Each contrastive block consists of two convolutional layers with ReLU activation, followed by 2x2 maxpooling. The block that operates on the lowest level has the same architecture, but does not do maxpooling. Each expansive block consists of a 2x2 upconvolution and the concatenation of the result with the output of the corresponding contrastive block. Then, two convolutional layers with ReLU activation are applied. Each contrastive block doubles the channel dimension, while each expansive block halves the channel dimension. Finally, the output of the U-net is cropped from the center such that the dimensionality of the output is the same as the input, to which it is summed (i.e., a residual connection).

    \item \textbf{Conditional Variational Autoencoder:}
    The CVAE consists of a prior network, generative network, and inference network. For the CVAE, the kernel size of the convolutional layers is 5x5 instead of 3x3. 
    \begin{itemize}
        \item Prior network $p_\theta(z | x^t)$: first, the image channel corresponding to the one-hot encoding of the walls is concatenated with the forward model's outputs. Then, two convolutional layers with ReLU activation are applied, followed by 2x2 maxpooling. Subsequently, two repetitions of a convolutional layer with ReLU activation followed by maxpooling are applied. Finally, the output is flattened and mapped to a lower-dimensional space using a linear layer with ReLU activation. Two separate linear layers map this output to the parameters of the normal distribution over the latent space $\mu$ and $\log \sigma$.
        \item Generative network $p_\theta(x^{t+1} |z, x^t)$: $z$ is concatenated channel-wise to the forward model's output, along with the wall channel of $x^t$. Two convolutional layers with ReLU activation are applied, followed by one convolutional layer with sigmoid activation, to obtain the pixel-wise Bernoulli probabilities of the cell's location. As the cell cannot be located on top of a wall, we explicitly set the decoder output probabilities for these pixels to 0.
        \item Inference network $q_\phi(z |x^{t+1}, x^t)$: the architecture of the inference network is identical to the prior network, with the exception that it takes an extra input channel, corresponding to the cell channel of the one-hot encoding of $x^{t+1}$. 
    \end{itemize}
\end{itemize}

\section{Hardware}\label{app:hardware}

Model training was done on a single Nvidia RTX 3060 GPU, and took at most three hours per model, depending on the dataset and training strategy. For a fair comparison of the simulation times of the Cellular Potts model and our method, we resorted to local hardware, as CompuCell3D did anyhow not support GPU acceleration for simulations without chemical diffusion. The CPU used was an Intel i7-9750H CPU clocked at 2.6GHz. For the GPU time measurements, we used an Nvidia Quadro P2000.

\section{Maximum Likelihood Sampling for Bifurcating Dynamics}\label{app:mlbif}
To quantitatively evaluate maximum likelihood sampling we again investigate the cell's position at the end of a simulation. Table~\ref{tab:bifur-hist-ml} contains the distributions of these end positions. While the model trained using the multi-step method approaches the ground-truth distribution, there are clear deviations. The volume-preserving model discussed in Section~\ref{sec:bifur-dyn-exp} much closer aligns with the Cellular Potts model and the expected distribution.
\begin{center}
\begin{tabular}{c|ccc}
Training method       & Left & Center & Right \\ \hline
One-step    & 0.04 & 0.35   & 0.61     \\
Pushforward  & 0.41 & 0.57   & 0.02  \\
Multi-step   & 0.34 & 0.53   & 0.13  \\ \hline
Cellular Potts       & 0.23 & 0.5   & 0.27  \\ \hline
Expected & 0.25 & 0.5    & 0.25 
\end{tabular}
\captionof{table}{Empirical distribution of the x-coordinate of the cell's center of mass at the end of the simulation. Trajectories are generated with the model that uses maximum likelihood sampling and various training strategies, and the Cellular Potts model.}
\label{tab:bifur-hist-ml}
\end{center}

\end{document}